\documentstyle[12pt,aasms4]{article}
\tighten

\newcommand\be{\begin{equation}}
\newcommand\ee{\end{equation}}

\input epsf
\lefthead{Basu and Antia}
\righthead{Solar flows: effect of peak-profile asymmetry}
\begin{document}

\title{Large scale flows in the solar interior: Effect of asymmetry
in peak profiles}

\author{Sarbani Basu}
\affil{Institute for Advanced Study, Olden Lane, Princeton NJ 08540,
U. S. A.}
\and
\author{H. M. Antia} 
\affil{Tata Institute of Fundamental Research, 
Homi Bhabha Road, Mumbai 400005, India}

\begin{abstract}
Ring diagram analysis can be used to study large
scale velocity fields in the outer part of the
solar convection zone. All previous works 
assume that the peak profiles in the solar oscillation power spectrum
are symmetric. However, it has now been demonstrated that the peaks are not
symmetric.
In this work we study how the explicit use of asymmetric peak profiles 
in ring-diagram analysis influences the estimated velocity fields.
 We find that the use of 
asymmetric profiles leads to significant
improvement in the fits, but the estimated velocity fields are not
substantially different from those obtained using a symmetric profile
to fit the peaks. The resulting velocity fields are compared with
those obtained by other investigators.
\end{abstract}


\section{Introduction}

Ring diagram analysis has been extensively used to infer horizontal
flows in outer part of the solar convection zone (Hill 1988;
Patr\'on et al.~1997; Basu, Antia \& Tripathy 1999).
This technique is based on the study of three-dimensional
(henceforth 3d)  power spectra of solar
p-modes on a part of the solar surface. In all the studies so far
the power spectra have been fitted to symmetric Lorentzian
peak profiles to calculate the frequency shifts due to velocity field.
It has been
demonstrated that in general, the peaks in solar
oscillation power spectra are not
symmetric (Duvall et al.~1993; Toutain 1993; Nigam \& Kosovichev 1998;
Toutain et al.~1998) and the use of symmetric profiles may cause
the fitted frequency to be shifted away from the true value.
The resulting frequency shift in the high degree modes using
ring diagram analysis has been found to be significant for f-modes
and low order p-modes (Antia \& Basu 1999).
Thus, it would be interesting to study how the use of asymmetric
profiles affects the measurement of flow velocities.

In this work we use the data obtained by the
Michelson Doppler Imager (MDI) on board the Solar and Heliospheric
Observatory  (SOHO) to measure the flow velocities in the
outer part of the solar convection zone. We fit the power spectra
using asymmetric peak profiles and compare the results with our
earlier results obtained using symmetric profiles (Basu et al.~1999).
 Although, it is possible to study the variation in
horizontal velocity with both longitude and latitude, in this work,
like in Basu et al.~(1999), 
we have only considered the longitudinal averages,
which contains information about the latitudinal variation
in these flows. For this purpose, at each latitude
we have summed the spectra obtained for different longitudes to get
an average spectrum which has information on the average flow velocity
at each latitude.
We have studied both the rotational and meridional components
of flow velocities.

The rest of the paper is organized as follows:  the
basic technique used to calculate the horizontal flow velocities using
ring diagrams is described in \S~2. The results are discussed in
\S~3, while 
the conclusions from our study are summarized in \S~4.

\section{The technique}

We adopt the ring diagram technique (Hill 1988; Patr\'on et al.~1997)
to obtain the 3d power spectra of solar oscillations.
We have used data from full-disk Dopplergrams obtained
by the MDI instrument of the Solar Oscillations Investigation (SOI)
on board SOHO. We have used the same spectra that were obtained
by Basu et al.~(1999). These spectra are based on regions covering
about $15^\circ\times15^\circ$ with $128\times128$ pixels
in heliographic longitude and latitude and centered at latitudes
ranging from $60^\circ$N to $60^\circ$S. Each region was tracked for
4096 minutes and we have taken averages over 12 spectra covering
an entire Carrington rotation in longitude. These spectra are spaced
by $30^\circ$ in longitude of the central meridian, which would correspond
to a time interval of about 3273 minutes. Thus there is some overlap between
different spectra. Similarly, we have taken spectra centered at
different latitudes separated by $5^\circ$ and hence there would be
considerable overlap in region covered by spectra centered at
neighboring latitudes.
The data cover the period from about May 24 to June 21, 1996.
Thus summing gives us a time averaged spectrum for the
period covered.

To extract the flow velocities and other mode parameters from the
3d power spectra we fit a model of the form
\be
P(k_x,k_y,\nu)=
{\exp(A_0+(k-k_0)A_1+A_2({k_x\over k})^2+
A_3{k_xk_y\over k^2})(S^2+(1+Sx)^2)\over
x^2+1}+{e^{B_1}\over k^3}+
{e^{B_2}\over k^4},
\ee
where
\be
x={\nu-ck^p-U_xk_x-U_yk_y\over w_0+w_1(k-k_0)},
\ee
$k^2=k_x^2+k_y^2$, $k$ being the total wave number,
and the 13 parameters $A_0, A_1, A_2, A_3, c, p,
U_x, U_y, w_0, w_1, S, B_1$ and $B_2$ are determined by fitting the spectra
using a maximum likelihood approach (Anderson, Duvall \& Jefferies~1990).
Here, $k_0$ is the central value of $k$ in the fitting interval and
$\exp(A_0)$ is the mean power in the ring.
 The coefficient
$A_1$ accounts for the variation in power with $k$ in the fitting interval,
while $A_2$ and $A_3$ terms account for the variation of power along the
ring. The term $ck^p$ is the mean frequency, while 
$U_xk_x$ and $U_yk_y$ represent the shift in
frequency due to large scale flows and the fitted values of $U_x$ and
$U_y$ give the average flow velocity over the region covered by the
power spectrum and the depth range where the corresponding mode is trapped.
The mean half-width is given by $w_0$, while $w_1$ takes care of
the variation in half-width with $k$ in the fitting interval.
The terms involving $B_1,B_2$ define the background
power, which is assumed to be of the same form as Patr\'on et al.~(1997).
$S$ is a parameter that controls the
asymmetry, and the form of asymmetry is the same as that
prescribed by Nigam \& Kosovichev~(1998).
This parameter is positive for positive asymmetry, i.e., more
power on the higher frequency side of the peak, and
negative for negative asymmetry. By setting $S=0$ we can fit 
symmetric Lorentzian profiles.

The fits are obtained by maximizing the likelihood
function $L$ or minimizing the function $F$ given by
\be
F=-\ln L=\sum_i\left(\ln M_i+{O_i\over M_i}\right),
\ee
where the  summation is taken over each pixel in the fitting interval. The term
$M_i$ is the result of evaluating the model given by Eq.~(1) at
$i^{\rm th}$ pixel defined by $k_x, k_y, \nu$ in the 3d power spectrum,
and $O_i$ is the observed power at the same pixel.
To evaluate the quality of the fit we use the merit function
(cf., Anderson et al.~1990)
\be
F_m=\sum_i\left(O_i-M_i\over M_i\right)^2
\ee
where the summation is over all pixels in the fitting interval.
Ideally, the merit function per degree of freedom should be close to unity.
The number of degrees of freedom can be defined as the difference between
the number of pixels in the fitting interval and the number of free
parameters in the model.

We fit each ring separately by using the portion of power spectrum
extending halfway to the adjoining rings. For each fit a region extending
about $\pm100\mu$Hz from the chosen central frequency is used.
We choose the central frequency for fit in the range of 2--5 mHz.
Power outside this range  is not significant. The
rings corresponding to $0\le n\le6$ have been fitted.
In this work we express $k$ in units of $R_\odot^{-1}$, which enables
us to identify it with the degree $\ell$ of the spherical harmonic of
the corresponding global mode.

The fitted $U_x$ and $U_y$ for each mode represents an average of the
velocities in the $x$ and $y$ directions
over the entire region in horizontal extent and over the vertical
region where the mode is trapped.
We can invert the fitted $U_x$ (or $U_y$) for a set of modes
to infer the variation in
horizontal flow velocity $u_x$ (or $u_y$) with depth.
We use the Regularized Least
Squares (RLS) as well as  the Optimally Localized Averages (OLA)
techniques for inversion as outlined by Basu et al.~(1999).
The results obtained
by these two independent inversion techniques are compared to test the
reliability of inversion results.
For the purpose of inversion, the fitted
values of $U_x$ and $U_y$ are interpolated to the nearest integral
value of $k$ (in units of $R_\odot^{-1}$) and then the kernels
computed from a full solar model with corresponding value of degree
$\ell$ are used for inversion. Since the fitted modes are trapped in
outer region of the Sun, inversions are carried out for $r>0.97R_\odot$
only.

\section{Results}

Following the procedure outlined in Section~2 we fit the form given by Eq.~(1)
to  suitable regions of the  3d spectra. We use approximately 900 modes
covering $200\le\ell\le1100$ and $2000\le \nu\le 5000$ $\mu$Hz for
each spectrum.
Fig.~1 shows some of the fitted
quantities for the averaged spectrum obtained from the region
centered at the equator.  This figure can be compared with
Fig.~2 in Basu et al.~(1999). The asymmetry parameter $S$ is found to
be significant and has a negative value for all the modes. The
magnitude of $S$ appears to increase with frequency.
However, the use of asymmetric profiles does not appear
to affect the fitted half-width or the horizontal
velocities $U_x$ and $U_y$ significantly. As with symmetric
profiles, the fitted half-width
$w_0$ appears to increase at low frequencies. This increase is probably
artificial since the actual width is less than the resolution limit of the
spectra. We have verified that keeping the width fixed during the fit
for these modes does not affect the values of $U_x$ and $U_y$ obtained
from the fits.

In order to test whether the fit
with additional parameter $S$ is indeed better we show in Fig.~2
the merit function [cf., Eq.~(4)] per degree of freedom
for fits to both symmetric and asymmetric profiles.
It is clear that the merit function has reduced significantly when
the additional  parameter $S$ is fitted. The value is close to unity
for all modes at low $n$ which can be identified by the ridges
in the low frequency end of Fig.~2.
The improvement in the fit is not very clear for higher order modes,
as the merit function does not appear to reduce
significantly. Similar results were found for fits to 2d spectra obtained
by averaging over the azimuthal direction (Antia \& Basu 1999).
In the 2d spectra too the f-modes were found to be distinctly asymmetric 
(Antia \& Basu 1999),
while the difference between fits to symmetric
and asymmetric profiles was progressively less clear for higher order modes, even though
the asymmetry parameter $S$ had similar magnitude.
This is mainly because the estimated errors increase with $n$ as the
corresponding rings in the 3d power spectra get smaller.
It may be noted that for symmetric profiles we have done a number of
experiments by including more terms denoting variations in parameters
not included in Eq.~(1), but these do not improve the fits significantly
and merit function does not reduce perceptibly (Basu et al. 1999). 
Thus it appears that
some asymmetry is indeed required to get good fits.
This asymmetry can be clearly seen in azimuthally averaged spectra
shown by Antia \& Basu (1999). The asymmetry parameter $S$ shown in
Fig.~1 is similar to what was found for the 2d fits in
Antia \& Basu (1999). Hence, it is clear that although it is difficult
to visualize asymmetry in 3d fits, the peak profiles are indeed
asymmetric.

There does not appear to be significant difference between the $U_x$
or $U_y$ obtained by fitting  symmetric or asymmetric profiles to
spectra obtained for the equatorial regions.
However, at high latitudes the
situation becomes somewhat different as can be seen from Figs.~3 and 4
which compare the fits to spectra from  regions
centered at a latitude of $50^\circ$N. It may be noted that even
for these latitudes the fitted values for the asymmetry parameter
$S$ are similar to what is found for equatorial spectra. However, the
fitted values of $U_x$ and $U_y$ are found to be more sensitive
to asymmetry at higher latitudes, probably due to effects of foreshortening.
The fits with the  asymmetric profile
give better results in general, as can be seen from the fact that  the ridges in $U_x,U_y$ for
different values of $n$ tend to merge better. The difference is
particularly significant in $U_y$. The steep trend in the different
ridges corresponding to low $n$ in  $U_y$ obtained by
fitting symmetric profiles
results in large number of outliers during the inversions. As a result,
we have to weed out a large number of modes to get reasonable inversion
results. 
We weed out all modes which have residuals larger than  $4\sigma$
for either $U_x$ or $U_y$ in an RLS inversion with low smoothing.
This does not happen when asymmetric profiles are used.
However, the fits are still not perfect at these latitudes even with
use of asymmetric profiles and the merit function per degree of
freedom is also somewhat
larger than unity for low order modes. Clearly, more terms or
a different form is required to improve the fits at high latitudes.

To study the effect of asymmetry on inverted
profiles for horizontal velocity components $u_x$ and $u_y$, we perform
inversions using both the fits and the results are shown in  Figs.~5 and 6.
It can be seen that except in deeper layers ($r\la0.98R_\odot$), the inverted results 
are similar regardless of whether we fit a symmetric or asymmetric 
profile to the spectra.
 From the
inversion results it is not possible to decide which fit is better,
but looking at the residuals and the fitted $U_x,U_y$ in Figs. 3,4
it appears that asymmetric profiles would give more reliable results.

The rotation velocity at each latitude can be
decomposed into the symmetric part [$(u_N+u_S)/2$] and an
antisymmetric part [$(u_N-u_S)/2$]. The symmetric part can be compared
with the rotation velocity as inferred from the splittings of global modes
(Schou et al.~1998) which sample just the symmetric part of the flow.
Since this component is not significantly affected by the asymmetry
in peak profile we do not show the detailed results.
However, the north-south antisymmetric
component of rotation velocity is small and its significance is not
well established from earlier studies and hence we reexamine these
results with fits to asymmetric peak profiles. The results for
near surface layers are shown in Fig.~7, which can be compared with
Fig.~7 of Basu et al.~(1999).
It can be seen that the results are not significantly different
from our earlier results.
In deeper layers the uncertainties
are larger and it is difficult to say anything about this component.

Since there is a distinct improvement in fits to $U_y$ at high
latitudes when asymmetric peak profiles are used,
we try to find its effect on meridional flow inferred
from ring diagram analysis.
Following Hathaway et al.~(1996) we try to
write the meridional component as
\be
u_y(r,\phi)=-\sum_i a_i(r) P_i^1(\cos(\phi))
\ee
where $\phi$ is the colatitude, and $P_i^1(x)$ are associated Legendre
polynomials. The first six terms in this expansion are found to be
significant and their
amplitudes are shown in Fig.~8, which can
be compared with Fig.~12 of Basu et al.~(1999). It can be seen that
use of asymmetric profiles introduces small changes in amplitudes
mainly in deeper layers. The amplitudes of all even components
increases slightly in deeper layers when asymmetric profiles are
used for fitting. In particular, the $P_4^1(\phi)$ component 
suggested by Durney (1993) is found to have an amplitude of
about 3--4 m/s for $0.97R_\odot<r<0.99R_\odot$.
There is no indication of any sign change in
the meridional flow up to the depth of $0.03R_\odot$, which
is consistent with results of Braun \& Fan (1998). However, the
magnitude of meridional flow velocity that we find (30 m/s) is much larger than
that obtained by Braun \& Fan, who find a mean meridional velocity
of 10--15 m/s averaged over latitudes $20^\circ$--$60^\circ$.
Clearly, more work is required to understand these differences.

We can also find the latitude at which $u_y=0$, which is the latitude near the  equator 
where the meridional flow diverges towards
the two poles.
The results shown in Fig.~9 can be compared with those obtained by
Gonz\'alez Hern\'andez et al.~(1999). In general we find this point
to be much closer to the equator as compared to what they have found.
The difference may be mostly due to the fact that we have averaged over all longitudes
in one Carrington rotation and as a result our error estimates are much
lower. For all depths we find that this point is within $4^\circ$ of the
equator and we do not find any chaotic behavior near the surface.
Of course, the resolution of inversions is limited to a depth below
$r=0.999R_\odot$ since we have only used modes with $\ell<1100$, which
have lower turning points below this depth. Thus the thin layer near the
surface is not resolved by inversions and we cannot expect reliable
results there. It is not clear if the small departure of this point
from the equator is significant as the difference in latitude estimated
from the two inversion techniques is comparable to the value.
A part of  the effect may also be due to
systematic errors arising from misalignment in the MDI instrument
(Giles et al.~1997) and  hence, no particular significance may be
ascribed to the location of this point.

\section{Conclusions}

Using the ring diagram technique applied to MDI data we have
determined horizontal velocities in the outer part of the solar
convection zone ($r>0.97R_\odot$).
We find that the use of asymmetric peak profiles improve the fits
to the 3d spectra ---- as measured by the merit function --- significantly.
The asymmetry parameter is found to be negative for all the modes,
i.e., there is more power on the lower frequency side of the peak.
This is the same as what has been found earlier at the low-degree
end of the power spectrum (Duvall et al.~1993; Toutain et al.~1998).
However, the use of asymmetric
profile does not affect the fitted velocities substantially.

The inferred meridional flow is dominated by the $\sin(2\theta)$ component
which has an amplitude of about 30 m/s in most of region covered in
this study. The $P_4^1(\phi)$ component suggested by Durney~(1993)
also has significant amplitude of about 3--4 m/s 
in deeper layers.
The north-south symmetric component of the meridional flow is
generally small and the meridional velocity (as a function of
latitude) changes sign at a
point very close to equator at all depths. This point marks the region
where flow diverges towards the two poles on two sides.
The small departure of this point from equator may not be significant.
There is no change in sign of meridional velocity with depth up to
about 21 Mm.

\acknowledgments

This work  utilizes data from the Solar Oscillations
Investigation / Michelson Doppler Imager (SOI/MDI) on the Solar
and Heliospheric Observatory (SOHO).  SOHO is a project of
international cooperation between ESA and NASA.
The authors would like to thank the SOI Science Support Center
and the SOI Ring Diagrams Team for assistance in data
processing. The data-processing modules used were
developed by Luiz A. Discher de Sa and Rick Bogart, with
contributions from Irene Gonz\'alez Hern\'andez and Peter Giles.

\vfill\eject

\begin{figure}
\plotone{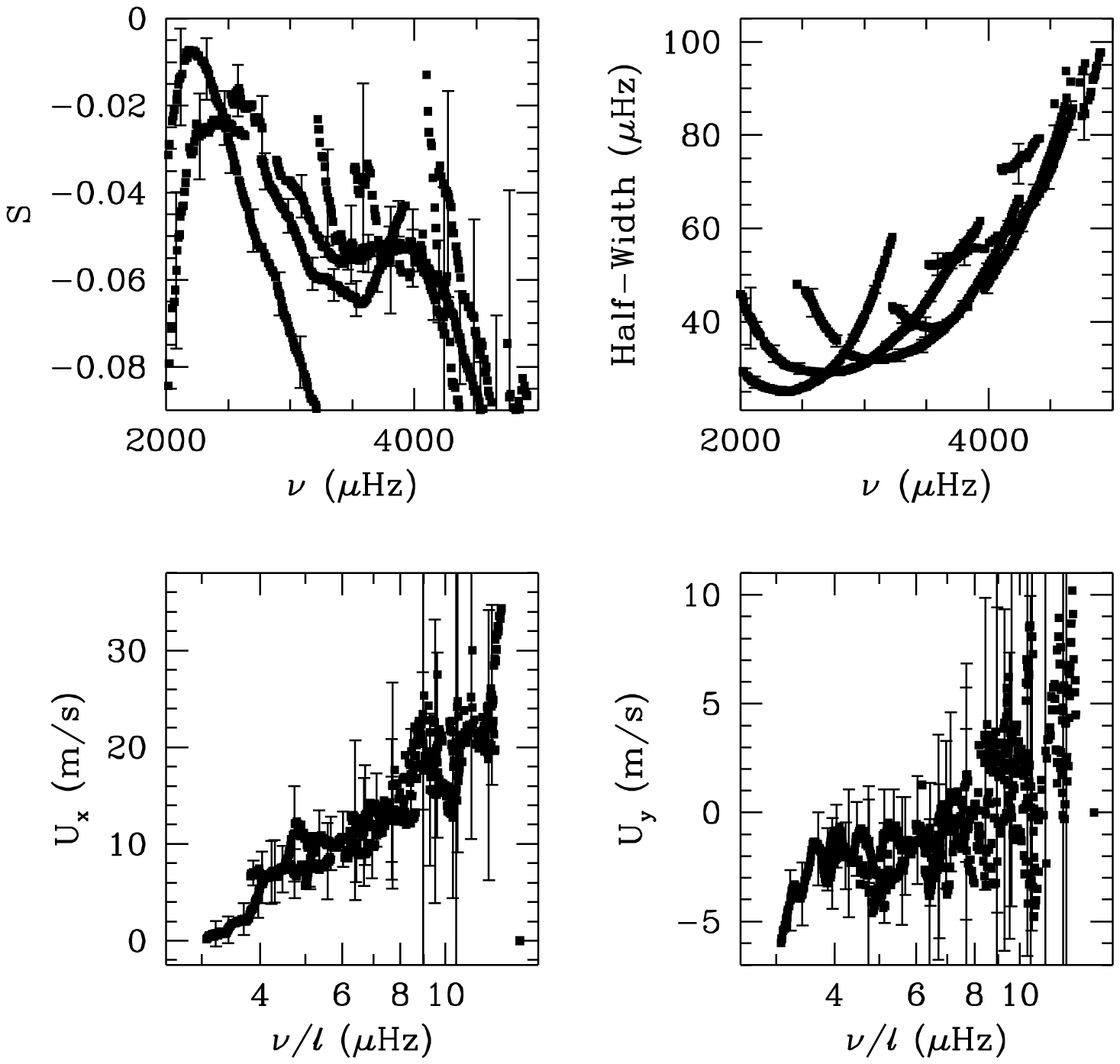}
\figcaption{The fitted parameters for the summed spectra centered at
the equator. This figure shows the asymmetry parameter $S$,
the half-width ($w_0$), and the average horizontal velocity $U_x, U_y$.
For clarity, only a few error bars are shown.}
\end{figure}

\begin{figure}
\plotone{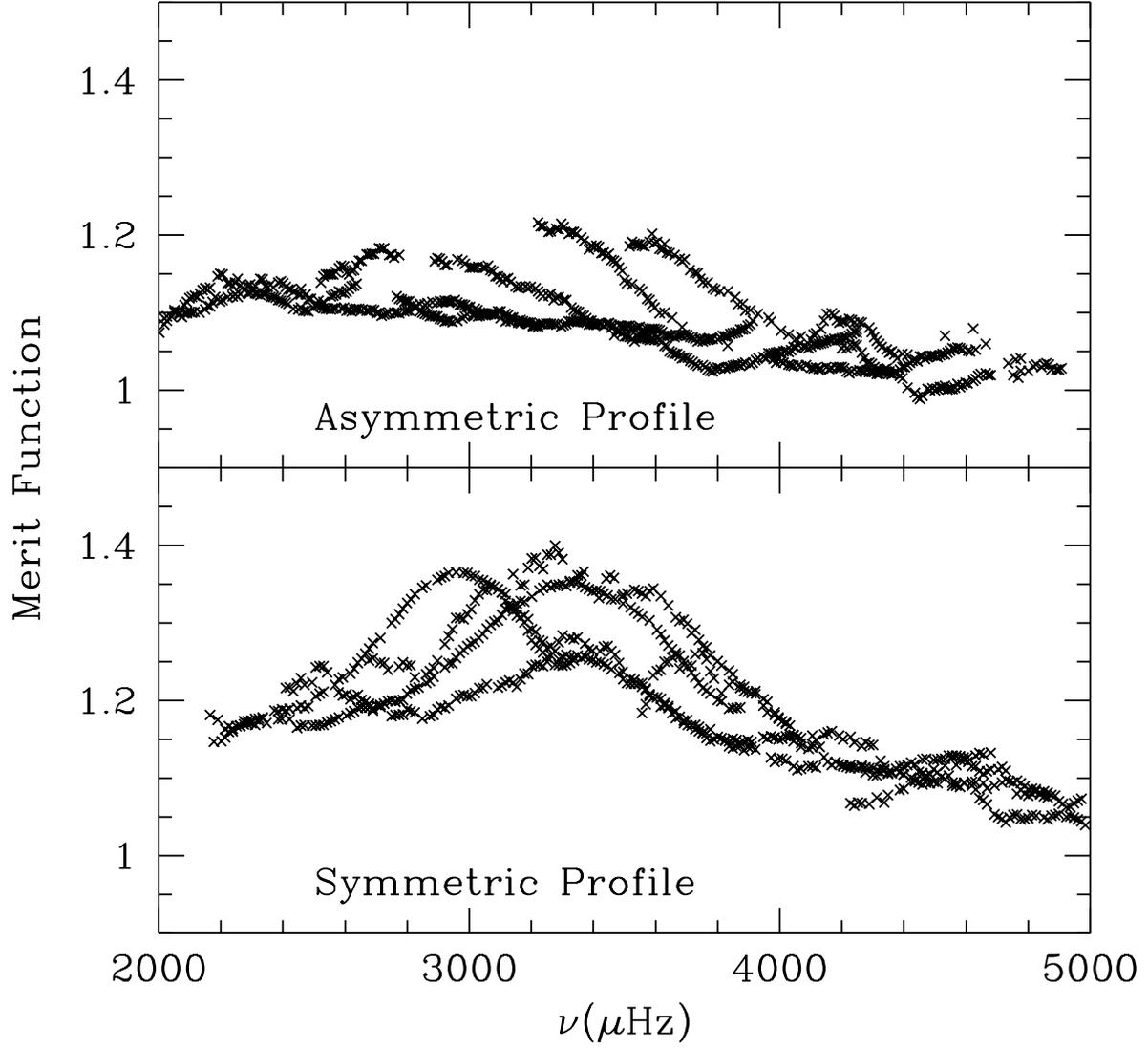}
\figcaption{The merit function per degree of freedom
for fits to the summed spectra centered at
the equator using symmetric and asymmetric profiles.
}
\end{figure}

\begin{figure}
\plotone{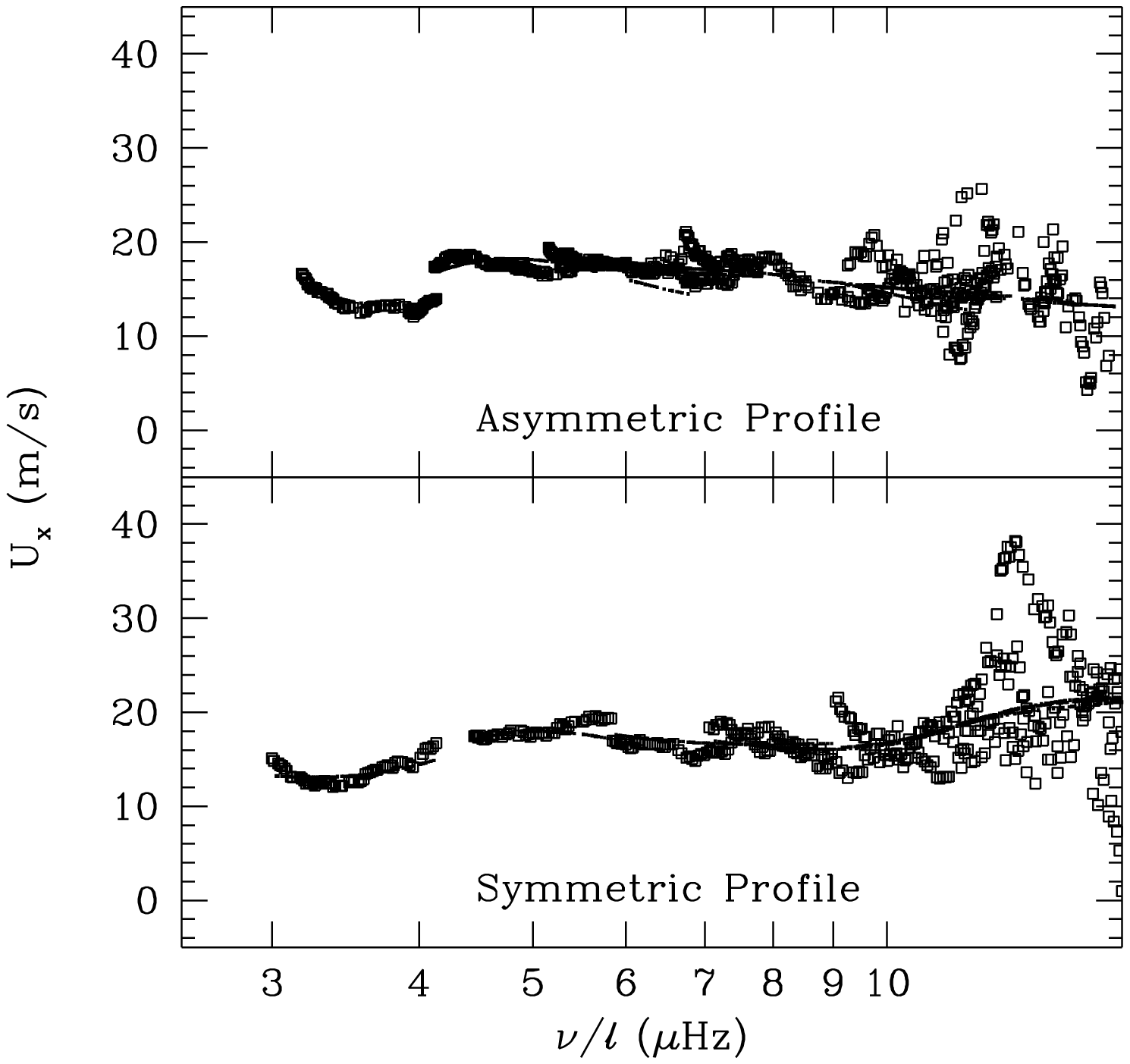}
\figcaption{The fitted horizontal velocity $U_x$ for fits to the
spectra centered at latitude of $50^\circ$N using symmetric and
asymmetric profiles. The open squares show the fitted $U_x$, while
the small filled squares which are merged into a band show the
calculated $U_x$ obtained from the (RLS) inverted velocity profiles shown
in Fig.~5.
}
\end{figure}

\begin{figure}
\plotone{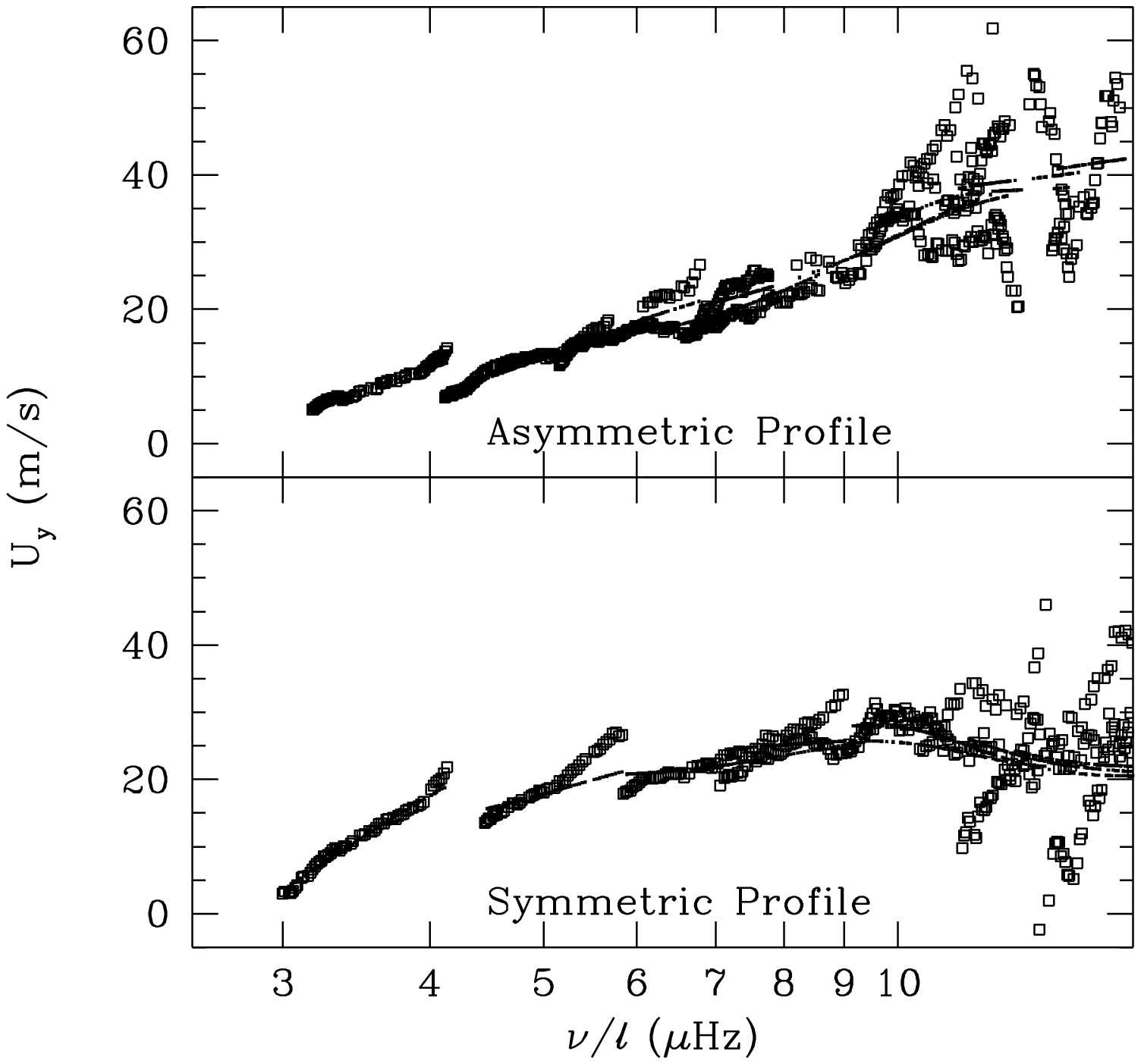}
\figcaption{The fitted horizontal velocity $U_y$ for fits to the 
spectra centered at latitude of $50^\circ$N using symmetric and
asymmetric profiles. The open squares show the fitted $U_y$, while
the small filled squares which are merged into a band show the
calculated $U_y$ obtained from the (RLS) inverted velocity profiles shown
in Fig.~6.
}
\end{figure}

\begin{figure}
\plotone{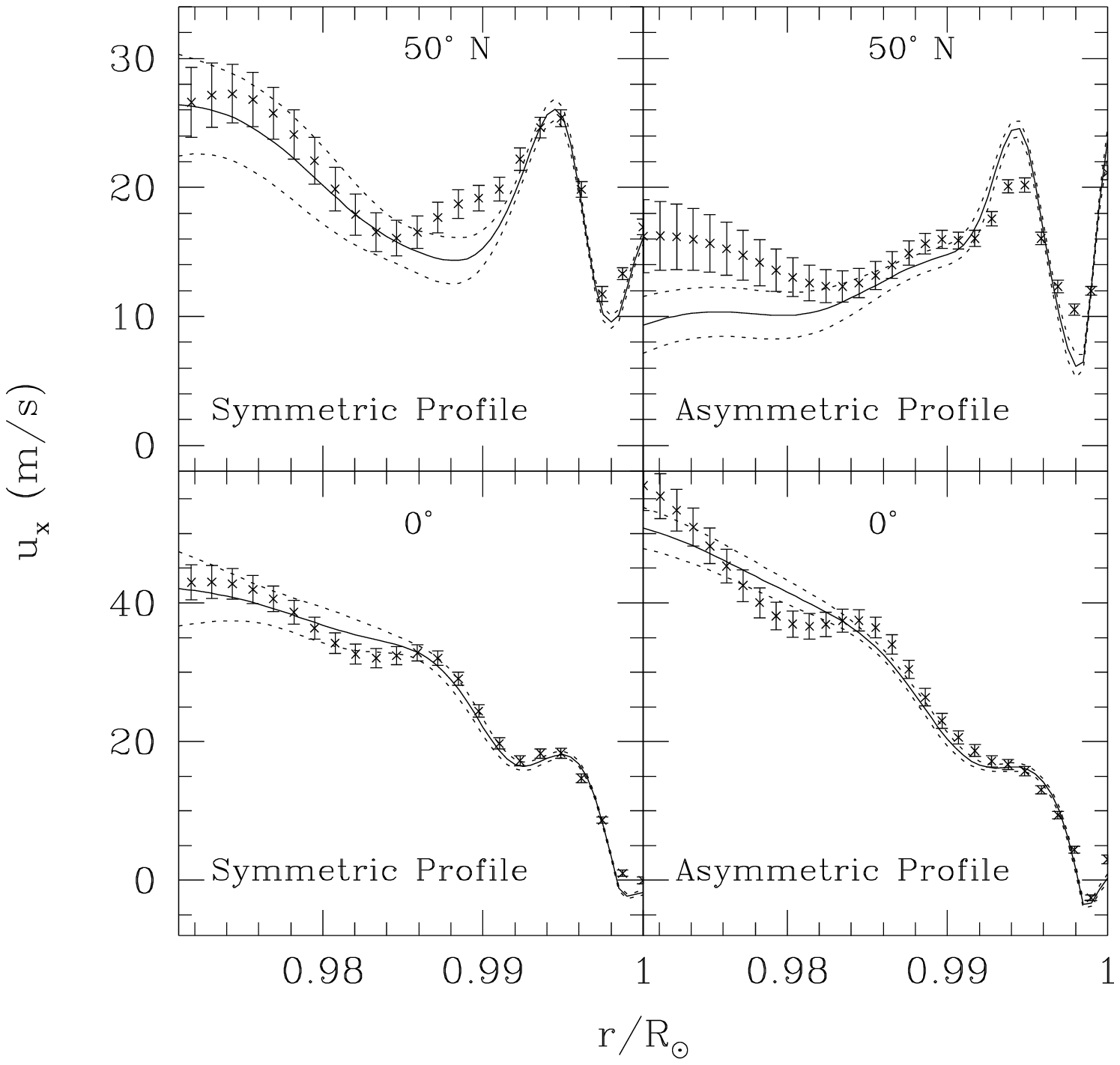}
\figcaption{The inverted horizontal velocity $u_x$ from fits to the
spectra centered at latitudes of $0^\circ$ and $50^\circ$N using
symmetric and asymmetric profiles. The continuous lines show the results
obtained using RLS with dotted lines representing the error limits, while
the points with error bars show the results using OLA technique.}
\end{figure}

\begin{figure}
\plotone{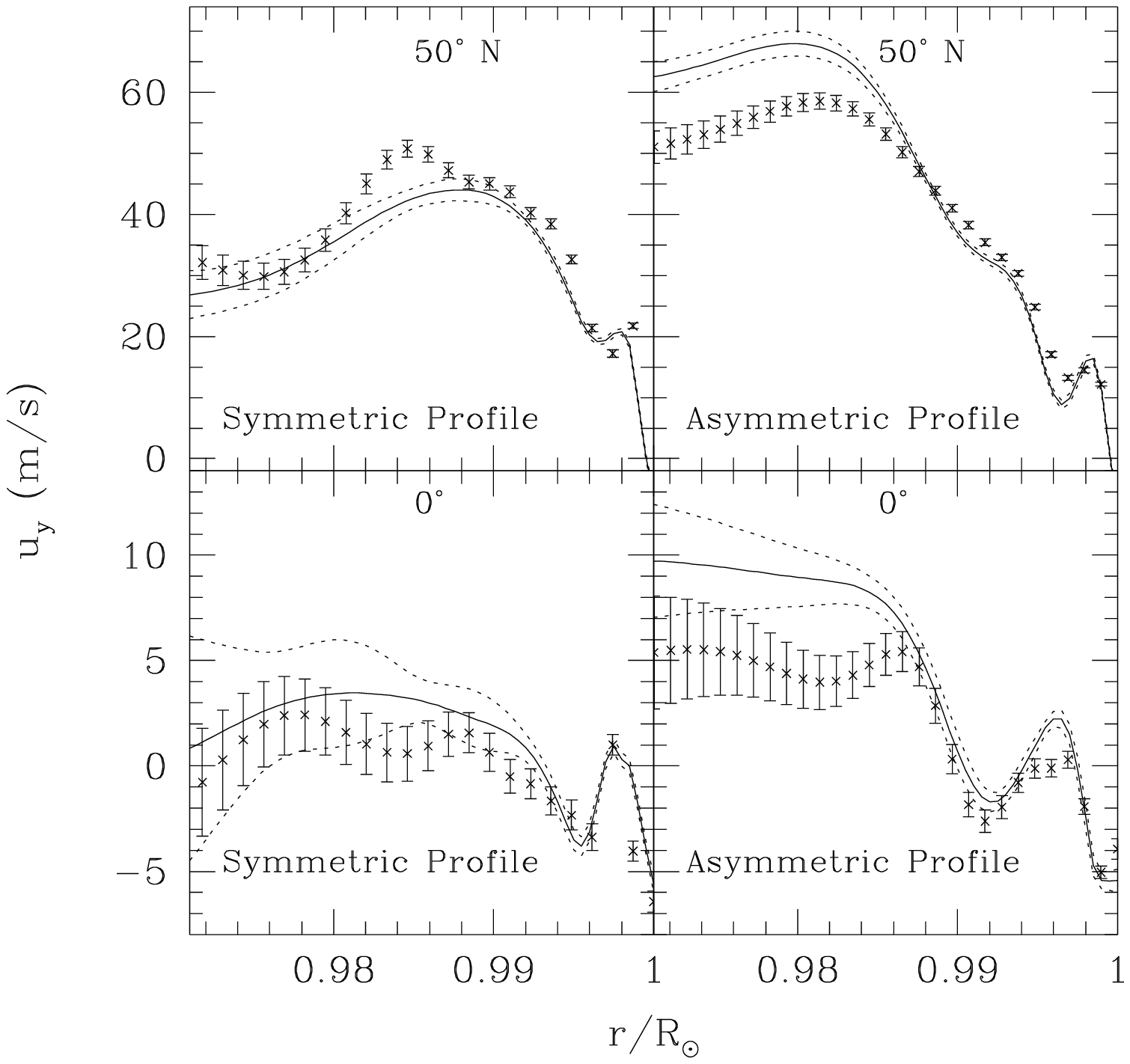}
\figcaption{The inverted horizontal velocity $u_y$ from fits to the
spectra centered at latitudes of $0^\circ$ and $50^\circ$N using
symmetric and asymmetric profiles. The continuous lines show the results
obtained using RLS with dotted lines representing the error limits, while
the points with error bars show the results using OLA technique.}
\end{figure}

\begin{figure}
\plotone{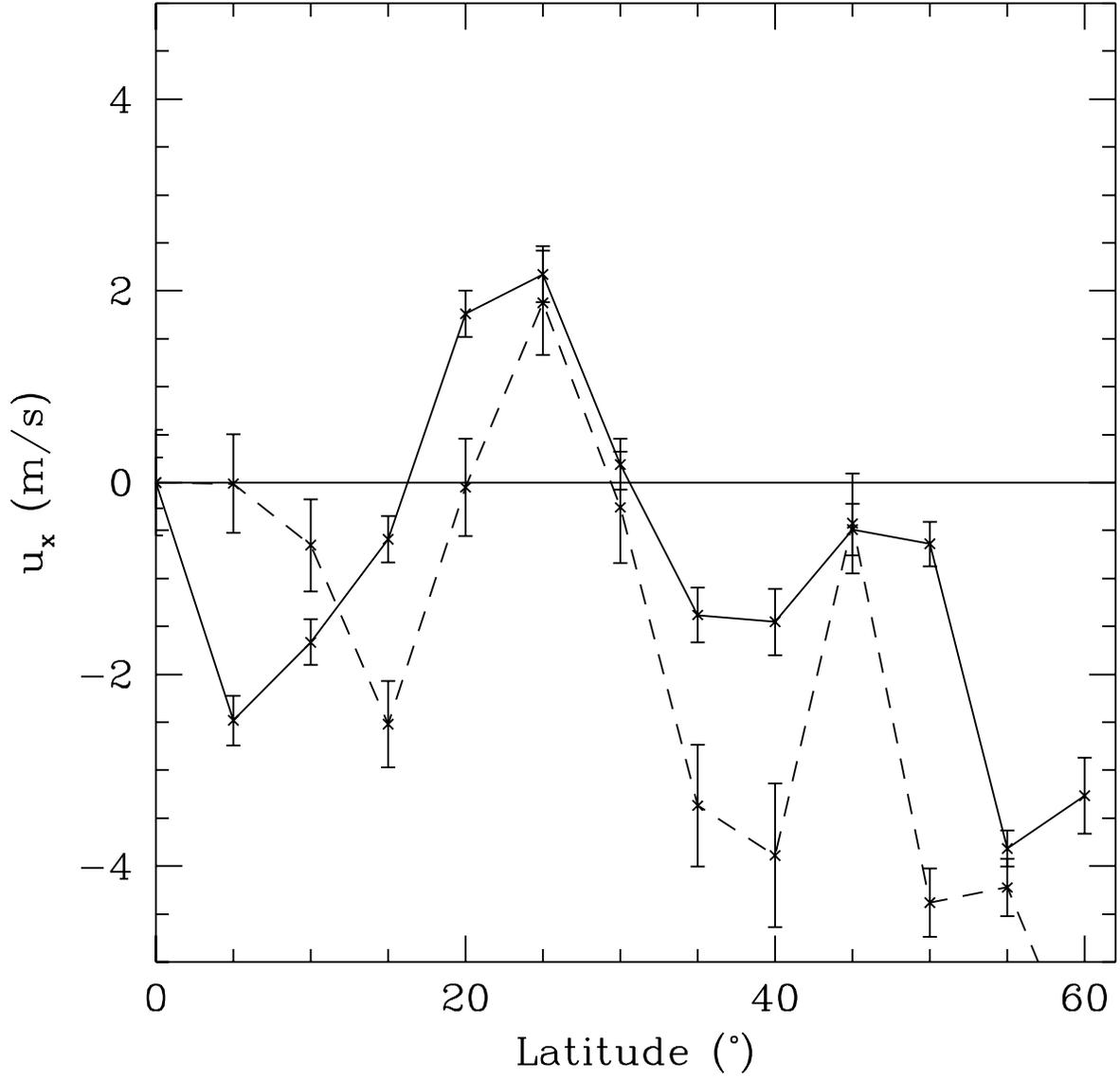}
\figcaption{The antisymmetric component [$(u_N-u_S)/2$] of the
rotation velocity plotted as a function of latitude for
$r=0.997R_\odot$ (continuous line) and $r=0.990R_\odot$ (dashed line).
These results are obtained using OLA technique.}
\end{figure}

\begin{figure}
\plotone{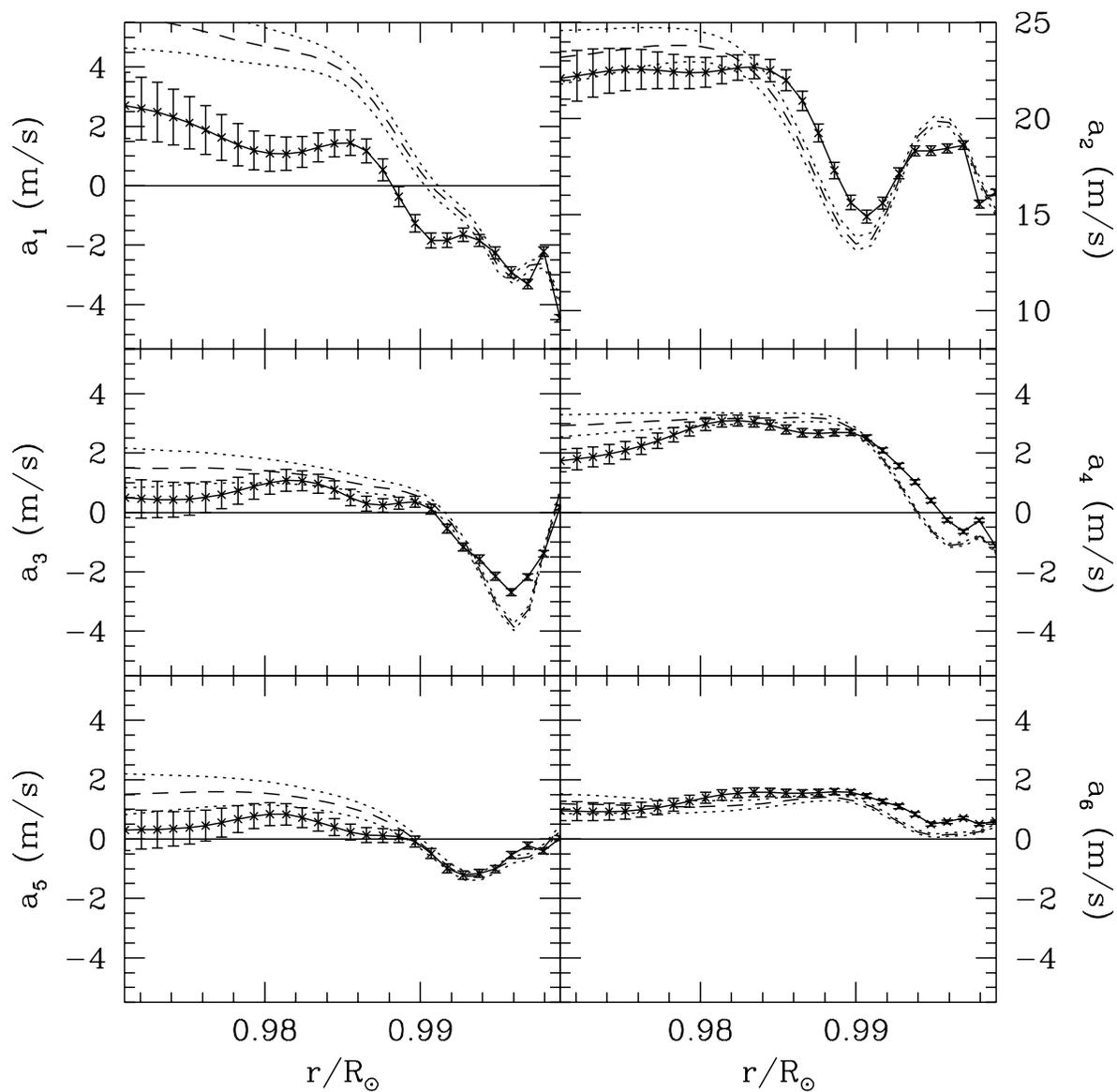}
\figcaption{Amplitude of various components of meridional velocity as a
function of depth obtained using fits to asymmetric peak profiles.
The continuous lines with error bars show the results
obtained using OLA inversions, while the dashed line shows the results
obtained using RLS technique with dotted lines showing the $1\sigma$
error limits.}
\end{figure}

\begin{figure}
\plotone{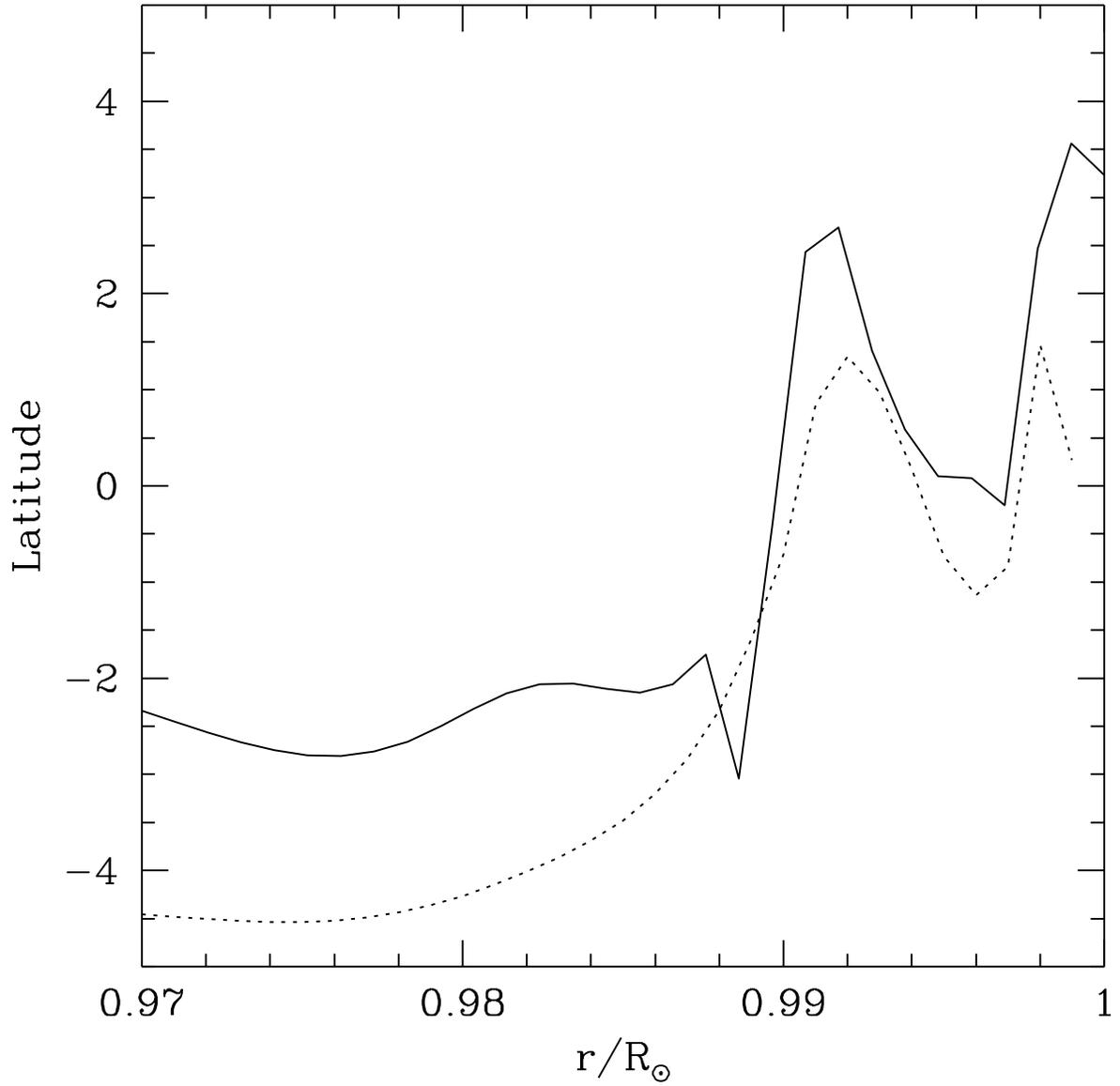}
\figcaption{Latitude at which $u_y=0$ is plotted as a function of depth
for results obtained using OLA (continuous line) and RLS (dashed line).}
\end{figure}

\end{document}